\documentclass[aip,jcp,nofootinbib,longbibliography]{revtex4-1}
\usepackage[utf8]{inputenc}
\usepackage{graphicx}
\usepackage{bm}
\usepackage{xcolor}
\usepackage{mathrsfs}
\usepackage{comment}
\usepackage{ulem}

\usepackage{ulem}

\usepackage[version=3]{mhchem}
\usepackage{siunitx}
\usepackage[pdftex, pdftitle={Article}, pdfauthor={Author}]{hyperref} 

\begin{document}
\title{From ultra-fast growth to avalanche growth in devitrifying glasses}
\author{Taiki Yanagishima}
\affiliation{Department of Physics, Graduate School of Science, Kyoto University, Kitashirakawa Oiwake-cho, Sakyo-ku, Kyoto 606-8502, Japan}
\email{yanagishima.taiki.8y@kyoto-u.ac.jp}
\author{John Russo}
\affiliation{Department of Physics, Sapienza University of Rome, P. le Aldo Moro 5, 00185 Rome, Italy}
\author{Roel P A Dullens}
\affiliation{Institute for Molecules and Materials, Radboud University Heyendaalseweg 135, 6525 AJ Nijmegen, The Netherlands}
\author{Hajime Tanaka}
\affiliation{Research Center for Advanced Science and Technology, The University of Tokyo, 4-6-1 Komaba, Meguro-ku, Tokyo 153-8904, Japan}
\affiliation{Department of Fundamental Engineering, Institute of Industrial Science, The University of Tokyo, 4-6-1 Komaba, Meguro-ku, Tokyo 153-8505, Japan}

\date{} 

\begin{abstract}
During devitrification, pre-existing crystallites grow by adding particles to their surface via a process that is either thermally activated (diffusive mode) or happens without kinetic barriers (fast crystal growth mode). It is yet unclear what factors determine the crystal growth mode and how to predict it. With simulations of repulsive hard-sphere-like (Weeks-Chandler-Andersen) glasses, we show for the first time that the same system at the same volume fraction and temperature can devitrify via both modes depending on the preparation protocol of the glass. We prepare two types of glass, a conventional glass (CG) via fast quenching and a uniform glass (UG) via density homogenization. Firstly, we bring either glass into contact with a crystal (X) and find the inherent structure (CGX/UGX). During energy minimization, the crystal front grows deep into the CG interface, while the growth is minimal for UG. When thermal noise is added, this behavior is reflected in different crystallization dynamics. CGX exhibits a density drop at the crystal growth front which correlates with enhanced dynamics at the interface and a fast growth mode. This mechanism may explain the faster crystal growth observed below the glass transition experimentally. In contrast, UGX grows via intermittent avalanche-like dynamics localized at the interface, a combination of localized mechanical defects and the exceptional mechanical stability imposed by the UG glass phase.
\end{abstract}

\maketitle

\section{Introduction}
Devitrification can be both a functional impediment for glassy materials \cite{Laitinen2013,Weng2020,Graeve2023} and a means to achieve tailored crystal growth \cite{Goncalves2002,Louzguine-Luzgin2014,Zhang2019}. Given the slow rate of diffusive transport in glasses and deeply supercooled melts, the mechanisms by which crystals can grow are largely different from what is expected from classical nucleation theory \cite{greet1967glass,Hikima1995,konishi2007possible,Sun2008,orava2014fast,powell2015fast,newman2020we,Lucas2023} and remains a field of active interest \cite{Sun2018,sun2020displacement,Gao2021,Hu2022}, but a complete mechanistic understanding of the process remains elusive.

It is known that the maturity, or the extent to which a glass is aged, strongly influences its structural evolution. In particular, hard-sphere and hard-sphere-like glasses have been shown to exhibit a transition from dynamically heterogeneous but continuous aging events \cite{Zaccarelli2009a,kawasaki2014structural} to intermittent, collective ``avalanche-like'' events \cite{Sanz2014}. The latter phenomena were not found to strongly correlate with metrics associated with the crystalline phase, like bond-orientational order \cite{Steinhardt1983,Lechner2008}, but rather with density heterogeneities frozen into the glassy structure during quenching \cite{Yanagishima2017}. The physical mechanism by which local density affects crystallization was shown to be mechanical as opposed to structural ordering related to crystallization, due to the close correlation between localized force chain evolution and particle dynamics.

Recently, it was found that the artificial elimination of global and microscopic density inhomogeneities could prevent aging events over significantly longer time scales than those over which avalanche-like dynamics is expected ~\cite{Yanagishima2021}. An algorithm was proposed to produce weakly polydisperse glassy states with anomalously high resistance to devitrification and aging through an iterative procedure that creates a uniform density field. Hereafter, we refer to glasses prepared via a conventional fast quench as conventional glasses (CG)~\cite{Yanagishima2017}, while those with a uniform density field as described in Ref.~\cite{Yanagishima2021} are referred to as uniform glasses (UG). The uniformity in the local mechanical environments was largely responsible for the nearly complete removal of non-affine displacements in UG glasses \cite{Yanagishima2021}. It was also found that post-critical nuclei frozen in the initial state did not experience any growth. This led us to consider the intriguing question of what happens when a deeply supercooled glass is placed in contact with a crystal (X). In this work, we reveal that CG and UG states give rise to two contrasting crystallization dynamics, ultra-fast growth and avalanche-like intermittent growth, respectively, at a finite temperature.

Fast growth refers to crystal growth whose kinetics are considerably faster than the time scales associated with mass transport in the glass. Recent work on a number of face-centered cubic (fcc) metals has revealed that fast growth is a consequence of the underlying crystalline order that characterized the inherent state, the local energy minimized configuration, of the crystal-liquid interface \cite{Sun2018}. This inherent state may be regarded as the wetting of a region of liquid structural order with crystal-like orientational order~\cite{tanaka2012bond} to the crystal surface~\cite{kawasaki2010formation,Hu2022}.

Here, we show that a CG state placed in contact with an fcc crystalline face followed by energy minimization (CGX) leads to a barrierless transformation of a large proportion of the glassy phase to an extended crystal phase up to what is expected from the equation of state at the pressure reached. In contrast, a UG state next to the same facet (UGX) does not show such behavior upon energy minimization, with any transformation limited to the interface because of the extraordinary mechanical stability of uniform glasses. When thermal noise is introduced, CGX and UGX again show distinct crystallization behaviors. CGX crystallizes with a fast growth mode due to a density drop at the crystal surface and the resulting enhancement of particle mobility at the growth front. In contrast, UGX crystallizes via intermittent avalanche dynamics. Avalanche-like growth is characterized by intermittent crystal growth ``events'' that are triggered by sudden bursts of collective particle displacements. In the context of deeply supercooled glasses, it has so far only been reported for spontaneous devitrification in bulk. To the best of our knowledge, the above is the first case of avalanche-like growth from a flat crystalline interface adjacent to a glass. Unlike bulk devitrification, we find interface-initiated dynamics due to the spatial coincidence of mechanical defects and the glass/crystal interface.

\section{Method}
\subsection{System and Numerical Methods}
The system we studied consisted of crystals and glasses of purely repulsive Weeks-Chandler Andersen (WCA) quasi-hard spheres in the volume fraction range $\phi = 0.479 \sim 0.532$. The volume fraction is defined as $\phi = \pi\sum{\sigma_i^3}/6V$, where $\sigma_i$ are the WCA diameters of all particles, and $V$ is the volume of the simulation box. Conventional glasses were generated using a modified Lubachaevsky-Stillinger (LS) algorithm in the same manner as in previous work\cite{Yanagishima2017}, featuring a rapid expansion of WCA spheres from low to high volume fraction while allowing for a limited number of particle displacements to relax frustration. Uniform glasses were generated from conventional glasses using the method described previously, where iterative particle size adjustments and FIRE energy minimizations were repeated to eliminate local density inhomogeneities \cite{Yanagishima2021}. Note that CG states are monodisperse, while UG states are slightly polydisperse; the effect of this will be discussed below.

We note that in the volume fraction range studied in this paper, $\phi = 0.479 \sim 0.532$, the glass transition temperature of a Lennard-Jones fluid is around $T_{\rm g,LJ} = 0.2 \sim 0.4$, according to simulation and theoretical works (summarized here~\cite{Robles2003}). In order to study glassy dynamics, Brownian Dynamics simulations were run at a temperature of $T = 0.025$, significantly below the glass transition point. We note that avalanche-mediated spontaneous aging has been confirmed for the same glass over this entire range\cite{Yanagishima2017}. All time scales are expressed in terms of the Brownian time $\tau_B = 3\pi\eta\langle\sigma\rangle^3/k_BT$, with unit viscosity $\eta$.

FIRE and steepest descent method minimizations were carried out using the LAMMPS software package. Devitrification dynamics at a finite temperature were simulated using Brownian Dynamics implemented in an openMP parallelized FORTRAN code.

\subsection{Structural characterisation}
To characterize local structure in our simulation, we primarily use three metrics. Firstly, the local volume fraction $\phi_i$ is found by taking a 3D Voronoi tessellation using the Voro++ package \cite{Rycroft2009}. Secondly, local bond-orientational order parameters are found to characterize crystallinity \cite{Steinhardt1983,Lechner2008}, specifically the $q_6$ parameter (averaged up to nearest neighbors only), commonly used to locate crystalline order in monodisperse hard-sphere systems. The complex vectorial form of the parameter is used to calculate bond coherence between particles $i$ and $j$ as $d_{6ij} = \bf{q_{6i}}\cdot\bf{q_{6j}}/|\bf{q_{6i}}||\bf{q_{6j}|}$. If this exceeds a threshold of 0.7, the bond is considered `solid' or `coherent'; the number of coherent nearest neighbors $n_X$ may be found for each particle. A particle with more than 6 solid bonds is considered crystalline in an fcc or hcp crystal \cite{Pusey2009}. Finally, the number of nearest neighbor particles in repulsive contact with a particle, i.e., within the range of the WCA potential, $n_{\rm FN}$ (or `force neighbors'), may be calculated to locate localized mechanical anomalies.

\subsection{Preparing adjacent glass/crystal sections}
\begin{figure}
    \centering
    \includegraphics[width=\columnwidth]{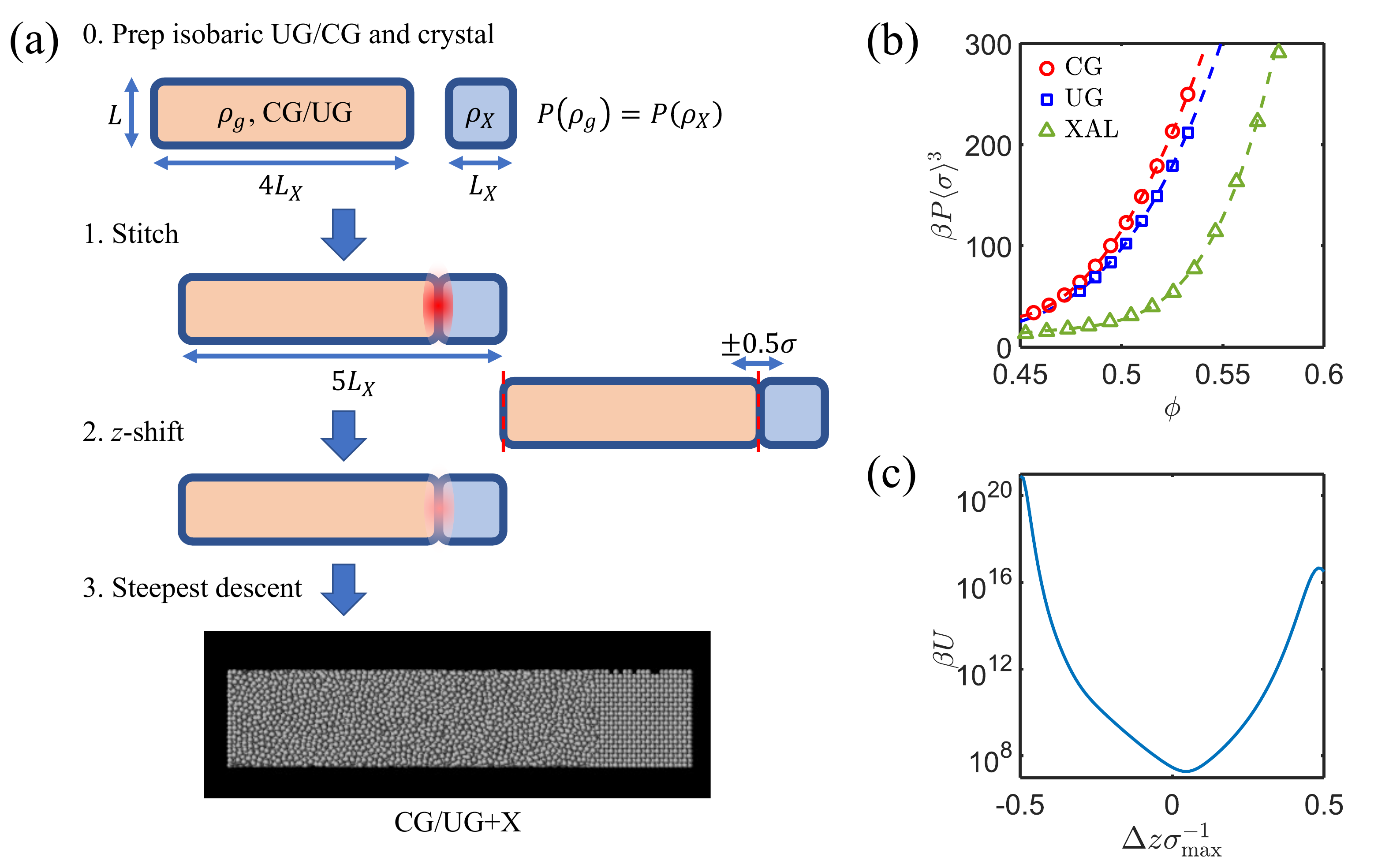}
    \caption{Generating isobaric inherent crystal/glass interfaces. (a) Schematic of the numerical protocol. (b) Equations of state. The normalized virial pressure as a function of local volume fraction $\phi$ are given for CG, UG, and crystal phases. $\beta = 1/k_{\rm B}T$ and $\langle\sigma\rangle$ is the average particle diameter. (c) Variation of total interaction energy $U$ with scaled adjustment, $\Delta z/\sigma_{\rm max}$, of glass portion positions in the $z$ direction, where $\sigma_{\rm max}$ is the largest particle diameter in the configuration.}
    \label{fig:schematic}
\end{figure}

To ensure mechanical stability when simulating a glass slab adjacent to a crystal, we must take into account the pressure of the two phases. At the same volume fraction $\phi$, a disordered glass has significantly higher pressure than a crystal; putting the two together in an isochoric ensemble would lead to a rapid expansion of the glass phase as the particles push out, and the crystal is compressed. Thus, we first prepared a set of conventional glasses, uniform glasses, and crystals by themselves at a range of volume fractions and measured the virial pressure. Thermal fluctuations were added, and the configurations were sampled over a single Brownian time (note that the vibrational fluctuations relax at significantly faster time scales). As shown in Fig.~\ref{fig:schematic}(b), the equation of state of the UG state is markedly different from that the CG state at the same volume fraction. These pressure profiles could be fitted with a polynomial to give an empirical variation of pressure with volume fraction. A 3rd order polynomial was used for the glassy phases, while a 4th order polynomial was used for the crystal.

To prepare a glass next to a crystal at a glass volume fraction $\phi_g$, we were able to locate a crystal volume fraction $\phi_X$ on the empirical fit which had the same pressure. A cubic box with $N_X = 4000$ particles in a face-centered cubic (fcc) configuration was resized to match $\phi_X$. Next, a completely separate cuboid box with the same $x$ and $y$ dimensions but a $z$ length $n$ times as long was prepared, and filled with the right number of particles to match the glass volume fraction $\phi_g$. CG glasses were prepared with $n=2$ and 4, while UG glasses were prepared with $n=4$ only, labeled $n$CG or $n$UG appropriately. A glass four times as long as the crystal was found to be large enough to allow investigation of the two growing crystal fronts on either side of the crystal without them interfering with each other, in most cases. Following this protocol, preparations at different $\phi_g$ lead to slightly different box sizes. These ranged between $L_0 = 15.462 \sim 15.814$ for high to low volume fraction.

Simply putting the two configurations next to each other created large particle overlaps at the edge of the box, which could not be relaxed with an energy minimization algorithm due to numerical considerations. Thus, the glass and crystal sections were firstly placed adjacent to each other to create a box with dimensions $\left(L_0,L_0,\left(1+4\right)L_0\right)$, then the particles in the glass section were displaced in the $z$-direction by $\Delta{z}$ with periodic boundary conditions to find a shift which minimized the energy due to particle overlap; the range of $\Delta{z}$ was kept to $\pm 0.5 \sigma_{\rm max}$, where $\sigma_{\rm max}$ is the diameter of the largest particle in the glass. All particles in the glass section were moved together by the same distance. This does not overtly bias the structure selection, but unambiguously locates a glass/crystal configuration with energy nearly 10 orders of magnitude smaller than with maximal overlap, as shown in Fig.~\ref{fig:schematic}(c). Once this configuration was obtained, a steepest descent algorithm was used to find the inherent structure. Note that the $z$-shift distinguishes this procedure from the scheme adopted by Tepper and Briels \cite{Tepper2001}, where a Lennard-Jones liquid was placed in contact with a crystal and allowed to relax via NVT dynamics for a limited number of steps. The result is an approximately isobaric glass/crystal interface at a local energy minimum. We refer to UG and CG states that are put adjacent to a crystal (X) in this way as UGX and CGX states.

\section{Results and Discussion}
\begin{figure}
    \centering
    \includegraphics[width=1.0\columnwidth]{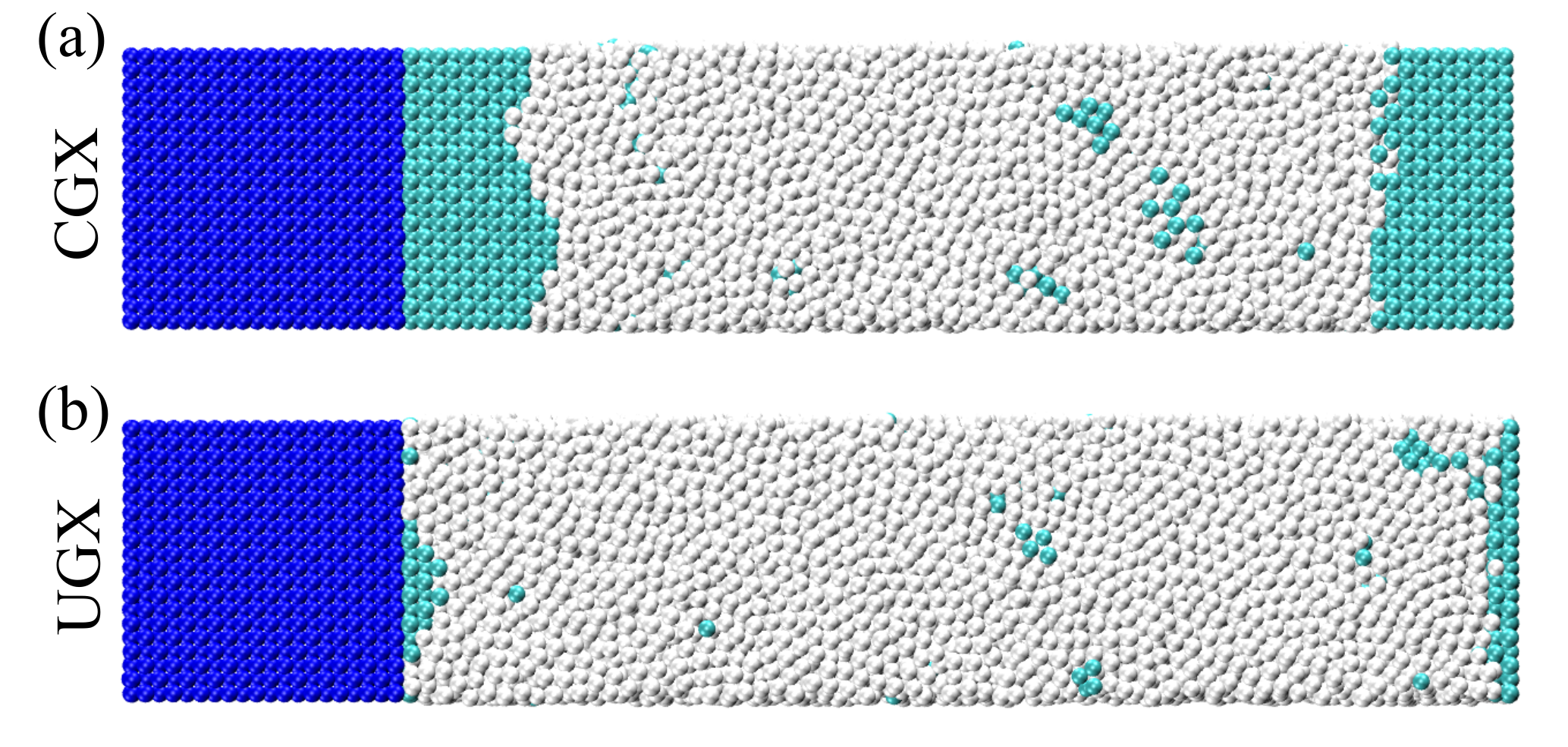}
    \caption{Inherent structures of (a) CGX and (b) UGX states at the same volume fraction of $\phi = 0.518$. Dark blue particles are the original crystalline section, light blue particles become crystalline on energy minimization, while white particles remain disordered.}
    \label{fig:CGvsUG}
\end{figure}

We start by considering the inherent states for CGX and UGX configurations generated by steepest descent minimizations as described above. The crystal is fcc, and the facet in contact with the glass is the (100) plane, unless otherwise stated. As shown in Fig.~\ref{fig:CGvsUG}, where we compare the inherent states obtained from the two glasses at the same volume fraction, we find drastically different results between them: while the CGX state shows a crystal that extends by significant distances into the glass bulk, the amorphous portion of the UGX state remains disordered.

\subsection{Inherent structure of CGX configurations}
\begin{figure}
    \centering
    \includegraphics[width=0.7\columnwidth]{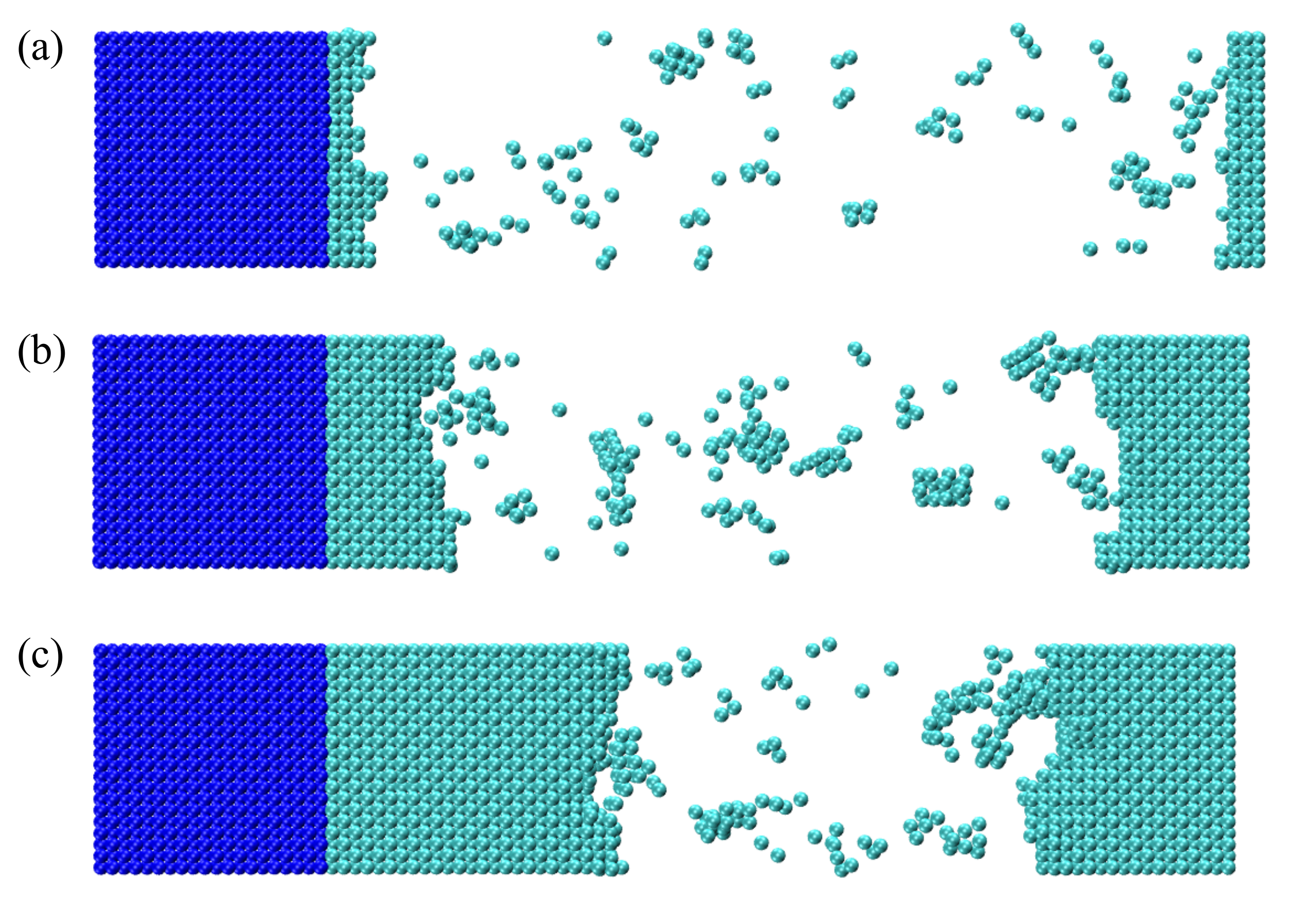}
    \caption{Inherent structures obtained from conventional glasses (CG) adjacent to a crystal (X). (a)-(c) correspond to inherent structures of isobaric crystal/glass configurations at increasing volume fractions, $\phi = 0.479$, 0.502, and 0.533. Dark blue particles are the original crystalline block; light blue particles are newly crystalline after energy minimization; all other particles are not shown.}
    \label{fig:CGstuff}
\end{figure}

Inherent structures for CGX states are shown for three different volume fractions in Fig.~\ref{fig:CGstuff}(a)-(c). It is clear that the crystal interface has significantly advanced for all volume fractions shown. This is reminiscent of the observation of Sun {\it et al.}\cite{sun2020displacement} that supercooled melts (above the glass transition) next to a crystal in metallic systems and Lennard-Jones fluids featured inherent structures with an advanced crystalline front, by up to several layers. Our findings show that this effect is far more pronounced for a glass, suggesting a different glass-specific mechanism. We also see that the advance of the interface is more pronounced at higher volume fractions. We may study this quantitatively by calculating the average value of $n_X$ in bins in the growth direction $z$ and finding where it becomes less than 6.5, since particles with $n_X \ge 7$ are considered crystalline. Within the range studied in this work, there is a monotonic relationship between glass density and the length $l_X$ by which the front advances, as shown in Fig.~\ref{fig:CGproperties}(a). This is reproduced with a smaller glass portion (2CGX) for low $\phi$: this is clearly not a finite-size effect.

Since the simulation box is of constant volume, the pressure inside the box drops as more particles become crystalline during energy minimization. This new, lower pressure may be used in conjunction with Fig.~\ref{fig:schematic}(b) and a lever rule to find what proportion of crystal and glass might be expected. This is multiplied by the box length and used to find an `expected' growth length, marked `Lever' in Fig.~\ref{fig:CGproperties}(a), averaged over the range of structures reached at different volume fractions. We find that this constitutes an upper bound to the length by which the crystalline front moves, a bound which is reached by the interface at higher volume fractions.

We note that one might think that this is triggered by protocol-specific features of the initial interfacial structure; despite the large drop in energy when the $z$-position of the glassy portion is shifted, interfacial energies are very large due to particle overlap. Indeed, extensive discussion exists in the literature for supercooled liquids, where sensitivity has been noted to the initial preparation of the structure \cite{Abraham1978,Broughton1982,Tepper2001}. However, we find that during steepest descent minimization, there is no significant structural evolution triggered by interfacial energy. In the context of deeply supercooled glasses, it seems that this is not what governs the advancement of the crystalline front. This is described in detail in the Supplementary Information.

A key question that arises is whether this advance in the crystalline front as a result of energy minimization is an intrinsic property of the CGX glass, the interface, or both. We will address this question in more depth later.

\subsection{Fast crystal growth from CGX configurations}
\begin{figure}[h]
    \centering
    \includegraphics[width=1.0\columnwidth]{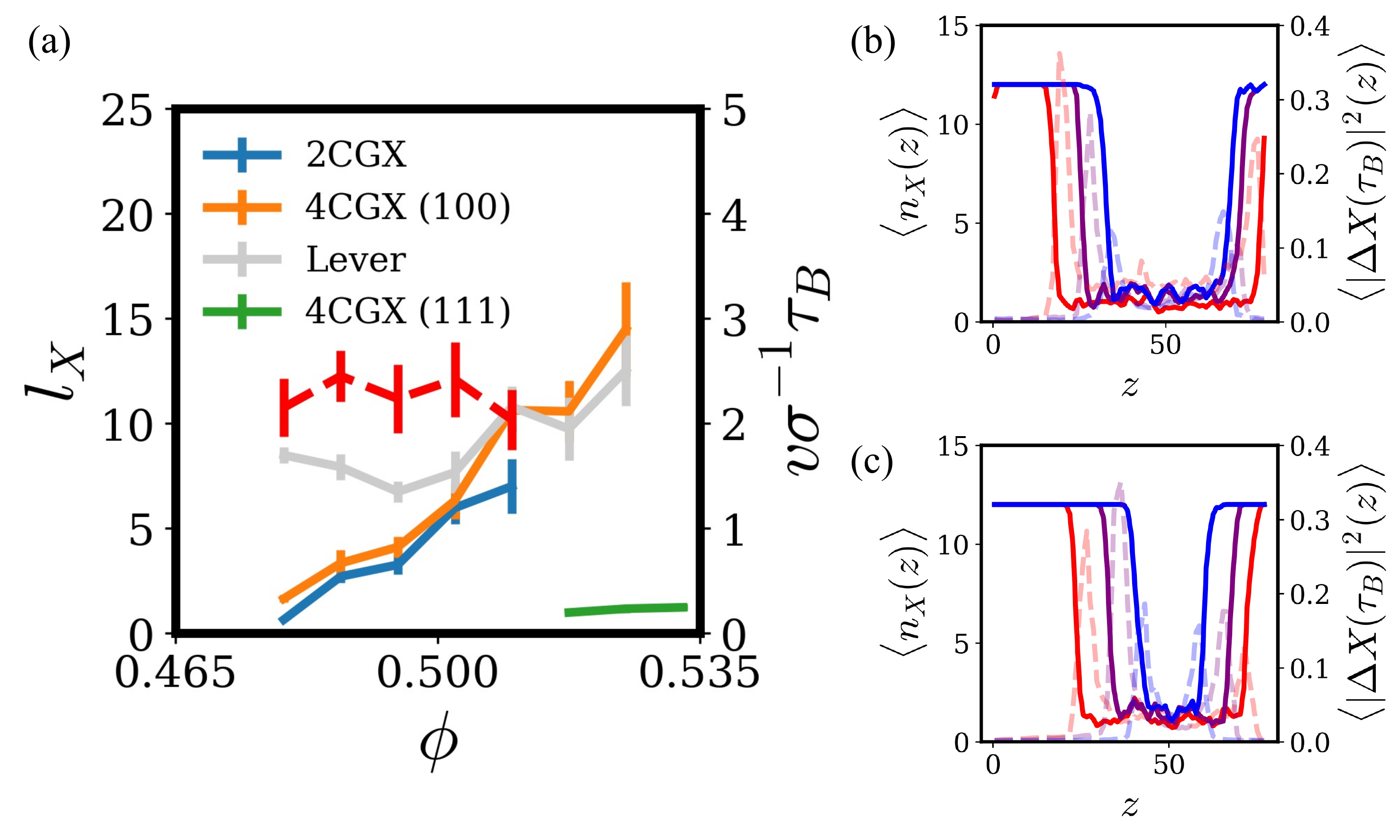}
    \caption{Inherent structure properties and crystallization dynamics of CGX states. (a) Length $l_X$ by which the crystalline front propagates during potential-energy minimization, found using two different lengths of glass section, 2CGX and 4CGX. The red dashed line shows the crystal front velocity $v$ via Brownian dynamics. (b) (solid lines) Average number of crystalline nearest neighbours $\langle n_X\rangle$ adopted by particles at different $z$ when $\phi = 0.479$ at $t = 0\tau_B$ (red), $4\tau_B$ (purple) and $8\tau_B$ (blue). (dashed lines) Average squared displacement of particles at different heights $z$ over $\Delta{t} = 1\tau_B$ from $t$. (c) The same as (b) for $\phi = 0.518$.}
    \label{fig:CGproperties}
\end{figure}

Next, the inherent structures are subjected to thermal noise and allowed to propagate over time via Brownian Dynamics. We find that the crystal front advances smoothly from both sides of the original crystalline block. The front can be tracked over time, as shown in Fig.~\ref{fig:CGproperties} for a trajectory where $\phi=0.475$ (b) and $\phi = 0.518$ (c). The front velocity $v$ is averaged over the first 1$\tau_B$ of the simulation and shown as a dotted line in Fig.~\ref{fig:CGproperties}(a) for all volume fractions studied. We stress that a conventionally prepared glass without an adjacent crystal at these volume fractions only features avalanche-like, intermittent dynamics, which lead to gradual devitrification over time. This velocity does not seem to be strongly volume fraction dependent.

We examine the average particle displacements at different $z$ positions with respect to the front. Unlike in avalanche dynamics, where dynamics are initiated by structural inhomogeneities in the bulk \cite{Yanagishima2017}, there is a clear, localized initiation of dynamics just in front of the crystalline front, shown in Fig.~\ref{fig:CGproperties}(b) and (c). We stress that all such events are initiated from local energy minima. Thus, despite all the initial states being mechanically stable initially, the barriers to configurations with advanced crystallization are very low.

\begin{figure}
    \centering
    \includegraphics[width=1.0\columnwidth]{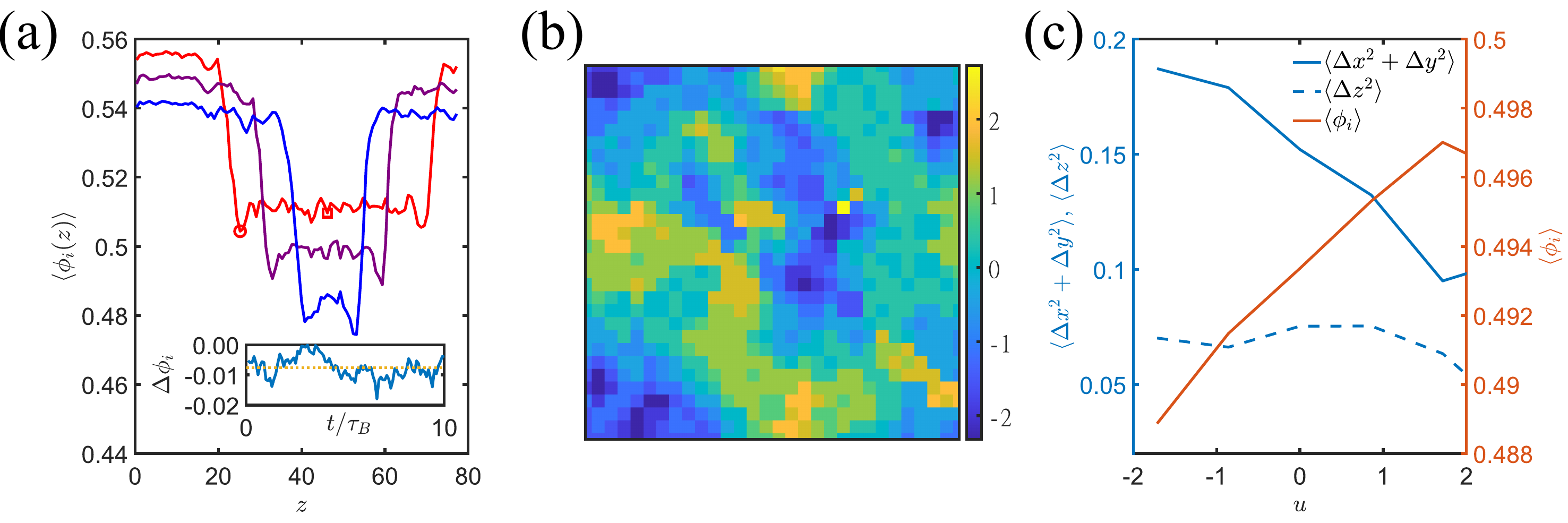}
    \caption{Interface structure of the growing crystal front in CGX configurations. (a) Local density profile of a growing front at 4$\tau_B$ intervals at $\phi = 0.518$. The profiles shift downward with time. There is always a dip at the interface. (inset) Depth of the dip over time. (b) Height $u(x,y)$ of the crystal-glass interface with respect to the mean. Color represents the depth of interface irregularities, in particle diameter units. (c) Average squared displacement sizes over $1\tau_B$ and density $\phi_i$ as a function of interface height $u$.}
    \label{fig:CGfront}
\end{figure}

We may study the structural characteristics of the growing front in more depth. Most notably, we find that there is a clear density anomaly localized to the interface, where a thin region has a density even lower than the glass. This is shown in Fig.~\ref{fig:CGfront}(a), looking at the density profile as a function of $z$ in a CGX configuration at $\phi = 0.518$ at $4\tau_B$ intervals like in Fig.~\ref{fig:CGproperties}(c). As the front moves, i.e., the box becomes more ordered, the volume fraction of both the crystal and glass drop simply due to the constant volume of the box, but the depth of the dip, as shown in the inset of Fig.~\ref{fig:CGfront}(a) is largely unchanged, as calculated from the difference between the minimum in the profile (circle in Fig.~\ref{fig:CGfront}) and the value in the middle of the originally glassy portion (square). We note that the presence of a `depletion zone' has been previously noted in both experiments~\cite{Sandomirski2014a} and simulations of Lennard-Jones fluids~\cite{Broughton1982,Huitema1999}, albeit at significantly less supercooling. An interfacial density drop has also been suggested as a potential factor in the ultra-fast growth of crystals in organic  glasses~\cite{tanaka2003possible,konishi2007possible,Yashima2022}.

We characterize this density anomaly further by considering the roughness of the crystalline interface. We firstly visualize the roughness by splitting the box into a cubic grid with a spacing of approximately 0.5$\sigma$, mapping the $n_X$ of the closest particle to each point; we may find how $n_X(z)$ evolves over thin sections in the $z$ direction and locate the position $z_{\rm int} (x,y)$, where $n_X(z)$ is no longer greater than 6. This is shown in Fig.~\ref{fig:CGfront}(b), where the color bar shows $u(x,y) = z_{\rm int}(x,y) - \langle z_{\rm int}\rangle$. A characteristic length of the roughness of around 5 particle diameters is consistent with the work of Gao {\it et al.} \cite{Gao2021}, which showed something similar in confocal experiments on colloidal suspensions; it also agrees with work by Dullens {\it et al.} \cite{Dullens2006}, where confocal experiments on colloidal particles sedimenting from a liquid phase onto a crystal showed similar fluid-crystal interfacial roughness at the extrapolated zero-gravity limit. Here, we go one step further and consider whether there is a more specific coupling between local density and/or particle dynamics with locations high or low within the roughness. Collecting statistics over a 10$\tau_B$ trajectory, we first find that the density drop is more acute the lower the interface is locally with respect to the mean interface position, i.e., in the `valleys' of the interface profile. Considering dynamics over 1$\tau_B$ intervals, the valleys are sites with the largest squared displacements parallel to the crystal face $\langle \Delta{x}^2 + \Delta{y}^2 \rangle$ (an `island' growth mode \cite{Gao2021}), though not $\langle \Delta{z}^2 \rangle$. Thus, for this growth mode, the largest displacements take place in the lowest-density regions localized to the depressions in the rough crystal-glass interface.

Here, we should note that in this study, we have adopted a scheme whereby the inherent structure is found, a $T = 0$ state, prior to Brownian Dynamics being activated at $T = 0.025$. This `detour' ensures that initial force imbalances at the interface are {\it not} the driving force behind crystal growth, just like interfacial energy is not responsible for the advancement of the crystal front during the energy minimization, as mentioned above. It may then be possible to interpret such initial force imbalances as a physical effect in itself and initiate the dynamics after the shift in $z$, albeit with an adaptive time step for the Brownian Dynamics simulation to account for the rapid relaxation of interfacial energy. This is beyond the scope of this work but would constitute interesting future work.

\subsection{UGX configurations and avalanche growth}
Uniform glasses placed adjacent to a crystal show drastically different behavior to CGX states. We focus on $\phi = 0.518$, the same volume fraction at which isolated UG states did not show any aging dynamics in previous work \cite{Yanagishima2021}. On observing the inherent structure of 4UGX, we immediately find that there is no advancement of the crystalline front, as shown in Fig.~\ref{fig:CGvsUG}. On measuring the shift of the position like for 4CGX, we find $l_X = 1.02 \pm 0.13$. There is a limited rearrangement of particles near the original crystal due to templating of the crystalline surface, but this does not lead to an advanced front.

\begin{figure}
\centering
\includegraphics[width=1.0\columnwidth]{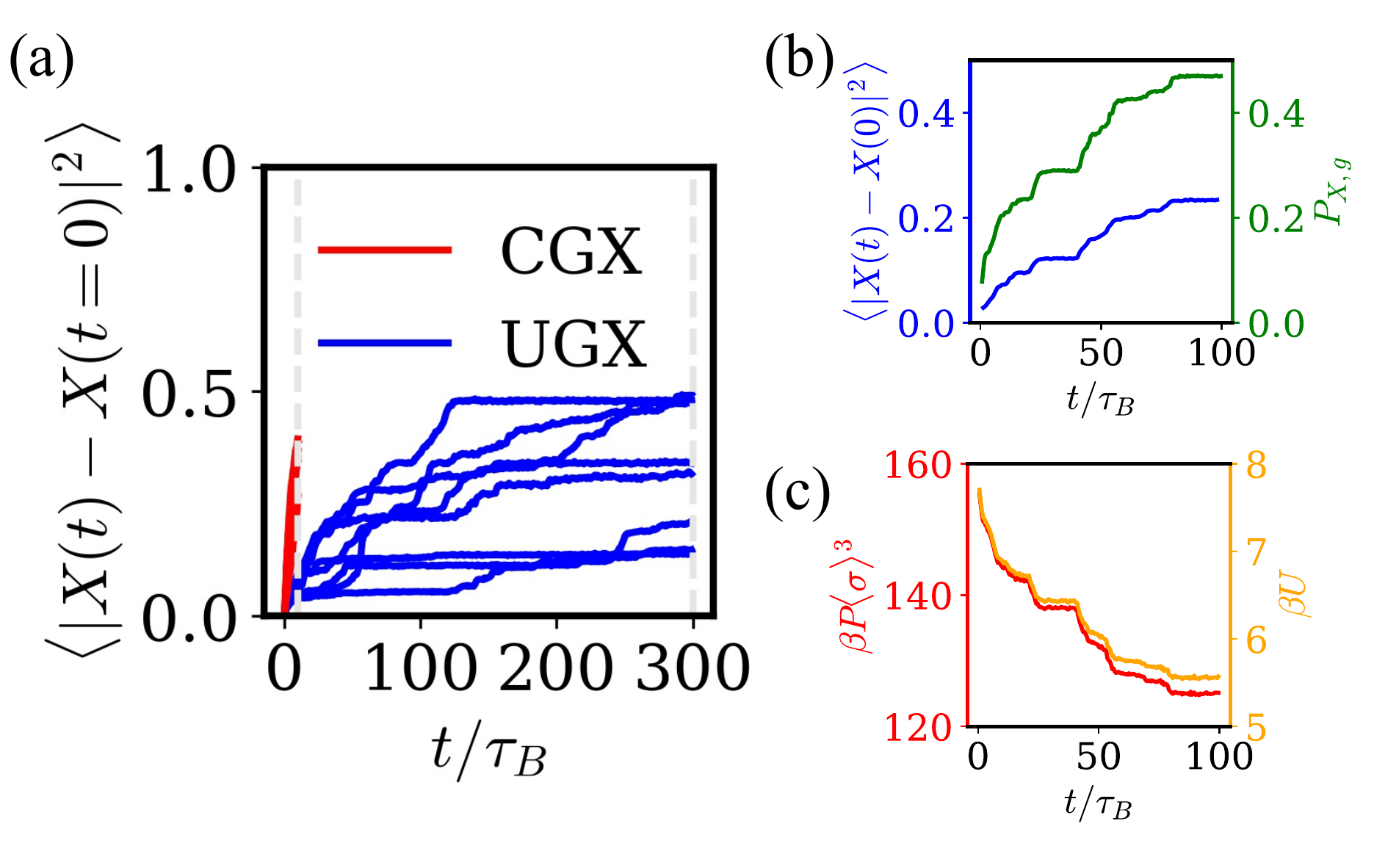}
\caption{Inherent structure properties and crystallization dynamics of UGX states at $\phi = 0.518$. (a) Squared displacement from the initial state due to Brownian Dynamics. Different lines indicate different simulation runs. Blue lines correspond to 4UGX states; red lines correspond to the dynamics of 4CGX states, for comparison. Dotted lines indicate simulation time. (b) Squared displacement and increase in the proportion of crystalline particles $P_{X,g}$ over time for an arbitrarily chosen run. (d) Normalized virial pressure $\beta P\langle\sigma\rangle^3$ and the total interaction energy $U$ normalized by the thermal energy $1/\beta$ over time.}
\label{fig:UGstuff}
\end{figure}

However, this does not mean that UGX states do not crystallize. We take these 4UGX states and subject them to Brownian Dynamics. Unlike the perfectly arrested devitrification seen in bulk UG states at the same volume fraction and temperature \cite{Yanagishima2021}, we find that there is an intermittent particle dynamics, as shown in Fig.~\ref{fig:UGstuff}(a). The displacement events are not always well separated like the avalanche events observed in bulk CG states, but they are nevertheless qualitatively different from the rapid crystallization of the 4CGX states, with individual events easily resolvable.

One of the trajectories is shown in detail in Fig.~\ref{fig:UGstuff}(b) and (c), where the proportion of crystalline particles in the originally glassy portion of the simulation box, $P_{X,g}$, the normalized virial pressure $\beta P\langle \sigma\rangle^3$ and the scaled energy $\beta U$ are also given. It is clear that these are correlated with each other, much like for avalanche dynamics of an ordinary, mature bulk glass \cite{Sanz2014}.

\begin{figure}[h]
\centering
\includegraphics[width=1.0\columnwidth]{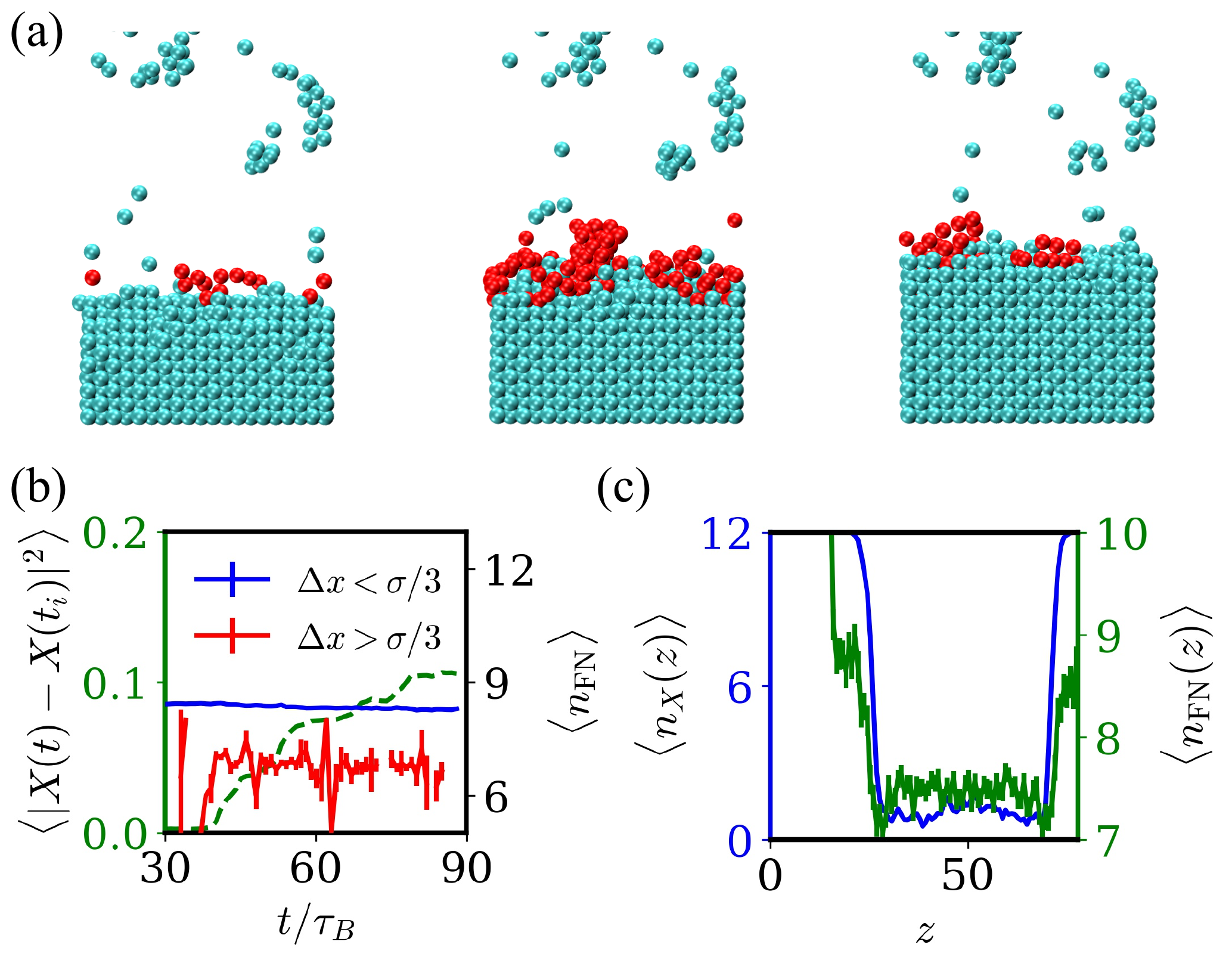}
\caption{Structural features of mobile particles during an intermittent crystallization event in a UGX state at $\phi = 0.518$. (a) From left to right, particles which are displaced more than $\langle\sigma\rangle/3$ in $1\tau_B$ (red) and crystalline particles (cyan) at 5$\tau_B$ intervals over the course of a crystallization event. (b) Average $n_{\rm FN}$ for mobile (red) and immobile particles (blue) over the course of an event, as judged from displacements over 1$\tau_B$ intervals. The dashed line shows the mean squared displacement from an arbitrary initial time $t_i$. (c) Profiles of the average numbers of crystalline nearest neighbors $\langle n_X \rangle$ and `force' neighbors $\langle n_{\rm FN} \rangle$ associated with particles at different $z$.}
\label{fig:UGevent}
\end{figure}

Given the discrete nature of the events, we may observe individual events and look for the origin of the displacement events as well as local structural precursors. This is shown in Fig.~\ref{fig:UGevent}(a), where particles displaced more than $\sigma/3$ in a 1$\tau_B$ interval are marked red; it is clear that the event is initiated from the interface, and subsequent particle motion is localized to the interface as the crystalline front grows. Though this is similar to the behavior of CGX, it is not consistent with the behavior of bulk CG systems in that avalanches are initiated from random mechanical inhomogeneities but not necessarily linked to the crystal interface~\cite{Yanagishima2021}. In the case of UGX, mechanical defects are localized near the crystal/glass interface. This can be clearly seen by looking at not only the density drop as in CGX, but the average number of nearest neighbors $n_{\rm FN}$ in mechanical contact at different positions with respect to the crystalline interface, as shown in Fig.~\ref{fig:UGevent}(c)).
A region near the interface has a slight depression in $n_{\rm FN}$ originating from the density and structural mismatch, which makes particles at the growth front mechanically less stable.

We also note that a direct link may be drawn between regions with low $n_{\rm FN}$ and particle dynamics, as shown by a time series where the average of $n_{\rm FN}$ for particles which are displaced by more than $\sigma/3$ in the interval $t_i+\tau_B$ are compared with that of particles which do not move. Particles that are able to take part in the event clearly have significantly fewer mechanical contacts. This trend continues throughout the event, validating the dynamic picture suggested above for a cascade of motion that follows the crystalline front as it grows, along with structural defects. Note that the UG {\it bulk} lacks such defects due to internal mechanical `homogenization' as described previously \cite{Yanagishima2021}.

We believe the spatial coincidence between the crystal interface and mechanical defects is largely responsible for the shorter time scale over which avalanches occur in this system compared to bulk CG systems. In bulk CG states, avalanche particles provide the perturbation for existing crystallites to grow. However, in UGX states, the close proximity of these sites to the crystal interface means that avalanche particles are localized at the interface. Thus, these particles directly enter into lattice sites with a domino-like cascade, which causes the front to grow. This leads to a scenario where a significant proportion of the particles that participate in the avalanche in fact become crystalline themselves.

An important question here is whether this might be a product of the polydispersity, or a specific property of UGX states. For the volume fraction above, we proceeded to look at the dynamics of reCGX states, i.e., CGX states made using a particle size distribution of 3.6\%, the same as an arbitrarily chosen UGX glass prepared at this density. We found that the devitrification dynamics was largely continuous and consistent with CGX states, albeit being slower, with a front migration speed of $v = 0.83 \pm 0.05$. The migration of the crystalline front during initial energy minimization is suppressed due to small lattice distortions presenting small barriers, but these do not entirely suppress a crystalline inherent structure at the front. The barriers are also easily surmountable. Details are given in the Supplementary Information.

We may approach all of these findings from the point of view of a potential/free energy landscape. In the case of CGX states, the original, inherent structure is a result of the original glass/crystal structure being destabilized by the introduction of a crystalline facet. The displacements required to accommodate this facet lead to a domino-like rearrangement, allowing the glass to transition to a more crystalline state. The length which this rearrangement can reach, $l_X$, is determined by the marginal mechanical stability of CGX states, which is the same property that causes spontaneous dynamics in bulk CG states to be intermittent \cite{Yanagishima2017}, and also by a pressure balance.

\begin{figure}
    \centering
    \includegraphics[width=10cm]{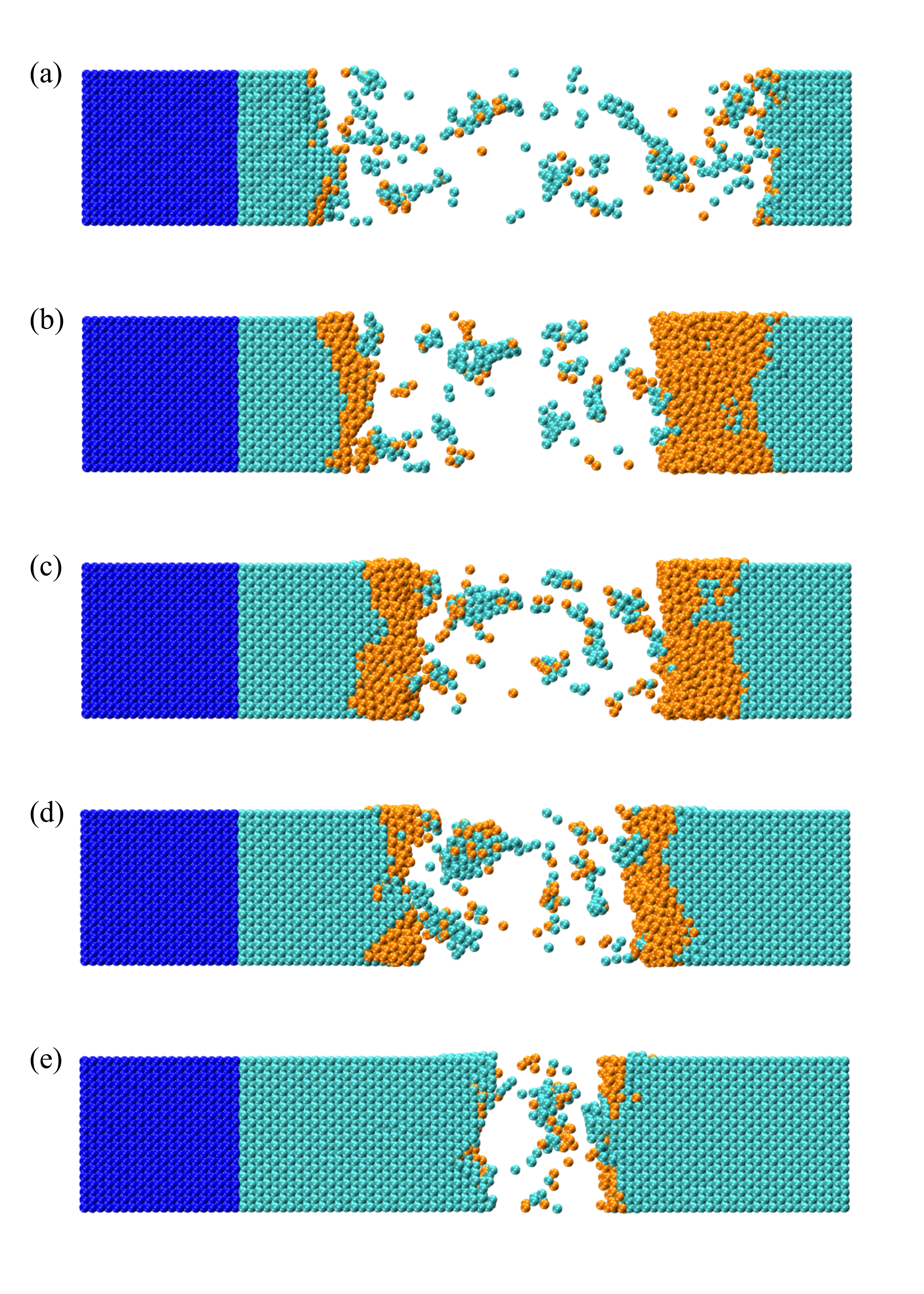}
    \caption{Crystalline portion of a growing CGX state at $\phi = 0.518$ at different times, along with the crystalline portion of the inherent structure. (a)-(e) correspond to $t =$~0.1, 1.0, 2.0, 4.0 and 8.0~$\tau_B$, respectively. Dark blue particles correspond to the original crystalline box, light blue particles correspond to newly crystalline particles at time $t$, and orange particles correspond to particles which are crystalline in the inherent structure. All colors are mapped onto particles at their positions prior to energy minimization.}
    \label{fig:DynInh}
\end{figure}

A more subtle question is whether such behavior is an intrinsic property of the respective glasses, the interface, or both. To explore this further, we look at the inherent structure of CGX interfaces {\it during the thermally-activated growth}. These are shown in Fig.~\ref{fig:DynInh}. Apart from panel (a) with a short period of thermalization, in which the configuration is trivially similar to its original inherent structure, the inherent structures for systems with longer thermalization gain a thick newly crystallized region. This region moves with the crystal interface before thermalization and shrinking; this behavior is equivalent to the decrease in $l_X$ with decreasing $\phi$, since the simulation box becomes more crystalline and $\phi$ of the glass phase also decreases, as shown in Fig.~\ref{fig:CGfront}(a). Thus, it seems that the length scale over which the crystal front advances is a property of the glass. This is also supported by the fact that only modifying the glass state to the UG state leads to the absence of crystal front advancement. This is due to the extraordinary mechanical stability of the UG force network \cite{Yanagishima2021} that can create a local energy minimum immediately in the vicinity of the original glass/crystal structure. Yet, even for the UGX, this is not enough to arrest crystallization entirely under thermal excitation. The barriers are low enough that dynamics initiated off localized mechanical instabilities at the interface can provide the perturbation required to advance the front, albeit in an intermittent way.

The contrasting inherent structure and dynamics for CGX and UGX states is clear evidence of how the mechanical stability of glasses affects structural evolution. However, this is not exclusively the case. Work by Broughton {\it et al.} \cite{Broughton1982,Burke1988} and Huitema {\it et al.} \cite{Huitema1999} found that the growth dynamics of Lennard-Jones crystals in a melt differed significantly when grown from (100) and (111) facets, with barrierless growth from the former and diffusion-limited activated growth from the latter. This was also highlighted by Sun {\it et al.}~\cite{Sun2018}, who found that crystal-melt interfaces of a supercooled Lennard-Jones fluid with a (111) face had an inherent structure with an advanced interface. We confirm similar results for our crystal-glass interfaces, with CGX interfaces showing no advancement of the interface during minimization when the crystal face was made hexagonal, i.e., (111) (see Fig.~\ref{fig:CGproperties}(a)). The difference may arise from the fact that the (111) face has multiple stacking possibilities, which frustrate crystal growth, whereas particle attachment on the (100) surface is unique and straightforward. Therefore, whether a crystal surface structure allows straightforward growth or not also matters.

\section{Conclusions}
In this work, we observe a wide range of growth dynamics for a crystal adjacent to a glass, depending on the preparation protocol of the glass. For conventionally prepared glasses in contact with a crystal (CGX), we find that the inherent structure has an extended crystalline front, whereas a uniform glass in the same setting (UGX) shows minimal extension due to its mechanical stability. The crystallinity of the inherent states is determined by a property of the bulk glass when put in contact with a crystalline surface. We stress that this is not the result of preordering in the vicinity of the interface, as the glass was put into contact without thermal relaxation. Moreover, the advancement of the crystalline front far exceeds any length scale associated with the interface. 

We also study the crystal growth of the CGX/UGX states at a finite temperature. For CGX, we have identified a close correlation between the biggest interface-localized density anomalies, the largest particle displacements parallel to the interface, and the lowest points in the featured interface roughness, pointing to the key role played by local density. This result provides a hint for understanding the sudden acceleration of crystal growth speed below the glass transition point in both experimental and numerical systems~\cite{greet1967glass,Hikima1995,konishi2007possible,Sun2008,orava2014fast,powell2015fast,newman2020we,Lucas2023}. We also find an underlying interfacial inherent structure during growth. Comparing UGX and CGX states, it seems that this may be as much an intrinsic property of the internal glass structure rather than being solely determined by density and temperature. For mechanically stable UGX states, we observed intermittent dynamics that was only seen previously in CG states in bulk. Unlike avalanches in CG states, however, we find that avalanche initiation occurs at the crystal/glass interface; the spatial coincidence between the only available mechanical defects in the system and the crystal leads to the crystallization of the UG states via an intermittent dynamics, again strongly underpinned by mechanical stability. These findings highlight the importance of mechanical stability for the stability of glasses against templated crystallization. Though we find vastly contrasted growth dynamics from CGX and UGX states, we also note that this selection is not exclusively determined by the glass. We confirm previous findings from equilibrated crystal-melt interfaces that find activated dynamics from a (111) fcc interface, but barrierless growth from (100), albeit in the absence of any thermal preordering at the interface. This implies that the properties of the crystal surface are also critical for fast crystal growth.

\section{Supplementary Material}
Supplemental Information has been provided as a separate PDF file, describing (a) the effect of the interfacial energy when crystal and glass configurations are placed adjacent to each other (prior to energy minimization), and (b) the effect of polydispersity, and whether it is responsible for the properties of UGX states. 

\section{Acknowledgements}
T.Y. would like to thank Prof. Jun Yamamoto for fruitful discussions. Computations were carried out at the Supercomputer Center, the Institute for Solid State Physics, the University of Tokyo, and the SuperComputer System, Institute for Chemical Research, Kyoto University. T.Y. also acknowledges a Kyoto University Internal Grant for Young Scientists (Start-up), a Toyota Riken Scholar Grant from the Toyota Physical and Chemical Research Institute, and the Japan Science and Technology Agency (JST) CREST Program Grant Number JPMJCR2095 for financial support. J.R. acknowledges support from the European Research Council Grant DLV-759187 and from ICSC – Centro Nazionale di Ricerca in High Performance Computing, Big Data and Quantum Computing, funded by the European Union – NextGenerationEU. R.P.A.D. acknowledges the support of a European Research Council Grant 724834-OMCIDC. H.T. acknowledges a Grant-in-aid for Specially Promoted Research (JP20H05619) from the Japan Society of the Promotion of Science (JSPS).

\section{Data availability}
The data that support the findings of this study are available from the corresponding author upon reasonable request.

\section{Author Contributions}
(Capitalized terms conform to CRediT) Conceptualization and Funding Acquisition was undertaken by all authors. T.Y. carried out Data Curation, Investigation, Methodology development and Visualization. Formal Analysis and Writing of the Original Draft was led by T.Y., and supported by J.R., R.P.A.D and H.T. Software was developed jointly by T.Y. and J.R. All authors participated equally in Writing - Review and Editing.

\bibliography{biblio}

\end{document}


\title{Supplementary Information for `From ultra-fast growth to avalanche growth in devitrifying glasses'}
\author{Taiki Yanagishima}
\affiliation{Department of Physics, Graduate School of Science, Kyoto University, Kitashirakawa Oiwake-cho, Sakyo-ku, Kyoto 606-8502, Japan}
\email{yanagishima.taiki.8y@kyoto-u.ac.jp}
\author{John Russo}
\affiliation{Department of Physics, Sapienza University of Rome, P. le Aldo Moro 5, 00185 Rome, Italy}
\author{Roel P A Dullens}
\affiliation{Institute for Molecules and Materials, Radboud University Heyendaalseweg 135, 6525 AJ Nijmegen, The Netherlands}
\author{Hajime Tanaka}
\affiliation{Research Center for Advanced Science and Technology, The University of Tokyo, 4-6-1 Komaba, Meguro-ku, Tokyo 153-8904, Japan}
\affiliation{Department of Fundamental Engineering, Institute of Industrial Science, The University of Tokyo, 4-6-1 Komaba, Meguro-ku, Tokyo 153-8505, Japan}

\date{} 

\maketitle

\section{Effect of interfacial energy on CGX generation}
During the generation of CGX states, boxes containing a crystal and glass at approximately equal virial pressures are placed adjacent to each other before particles in the glassy portion are moved together in the $z$-direction by $\pm 0.5 \sigma_{\rm max}$, where $\sigma_{\rm max}$ is the largest particle diameter in the glassy portion. Despite the significant energy drop this entails, there is still a considerable amount of interfacial energy due to particle-particle overlaps. This leads to the possibility that the state reached by the subsequent steepest descent minimization is affected by this energy.

\begin{figure}
    \centering
    \includegraphics[width=8cm]{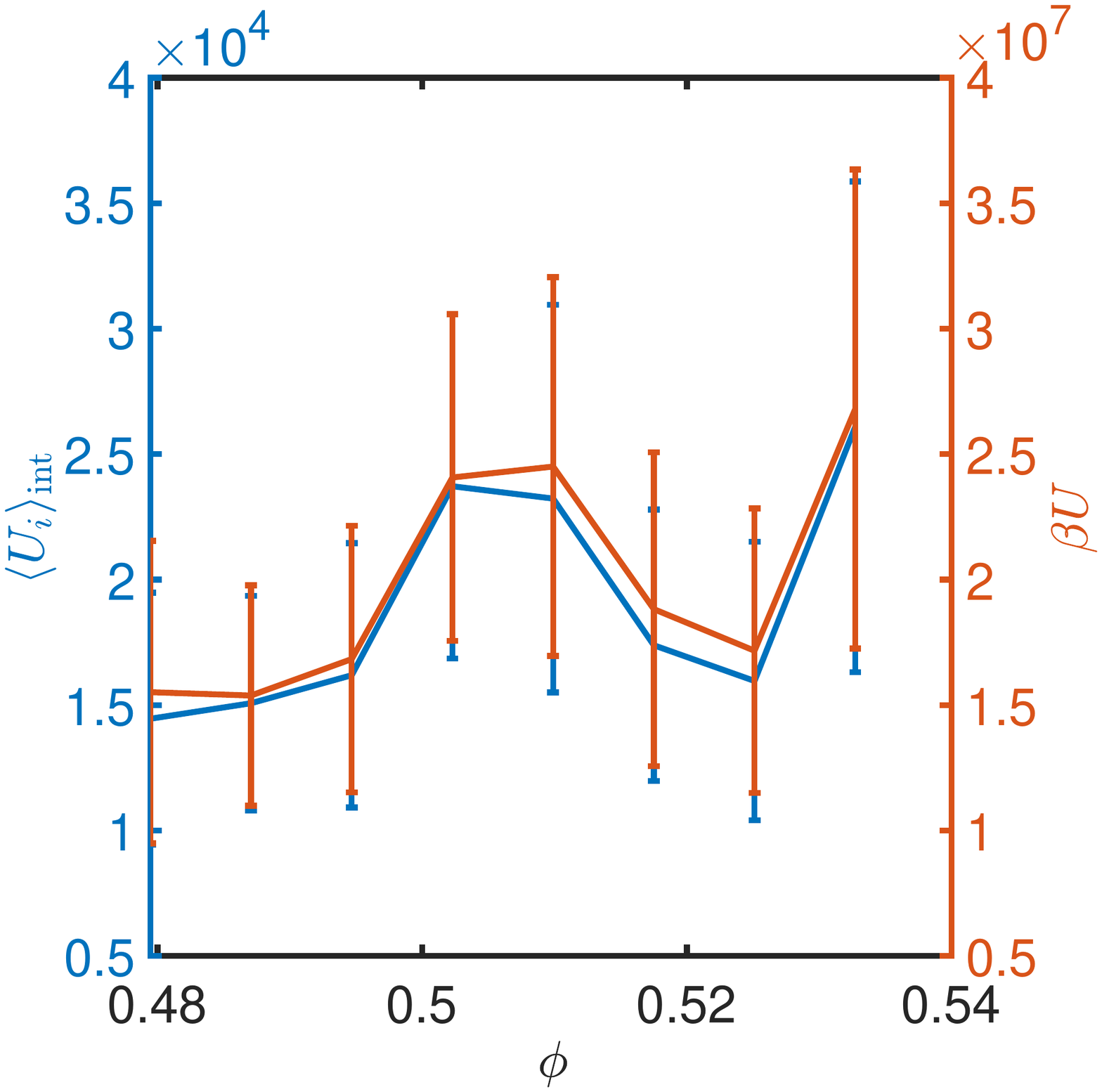}
    \caption{Average energy per particle of particles within two diameters of the interface $\beta\langle U_i\rangle_{\rm int}$, and the total energy $\beta U$ of configurations at different volume fractions $\phi$. All points are averaged over eight independent configurations; error bars are standard deviations over the configurations.}
    \label{fig:intU}
\end{figure}

On further investigation, it turns out this is not the case. Firstly, despite correlation between the distance moved by the crystalline front during minimization $l_X$ and the volume fraction $\phi$, there is no correlation with the average energy of particles near the interface, as well as the energy of the system as a whole. This is shown in Fig.\ref{fig:intU}, where there is no significant difference between states at different $\phi$. Note the close correlation between interfacial energy and the total energy; the interfacial energy is very large, and constitutes the majority of internal energy in the system.

\begin{figure}
    \centering\includegraphics[width=1.0\columnwidth]{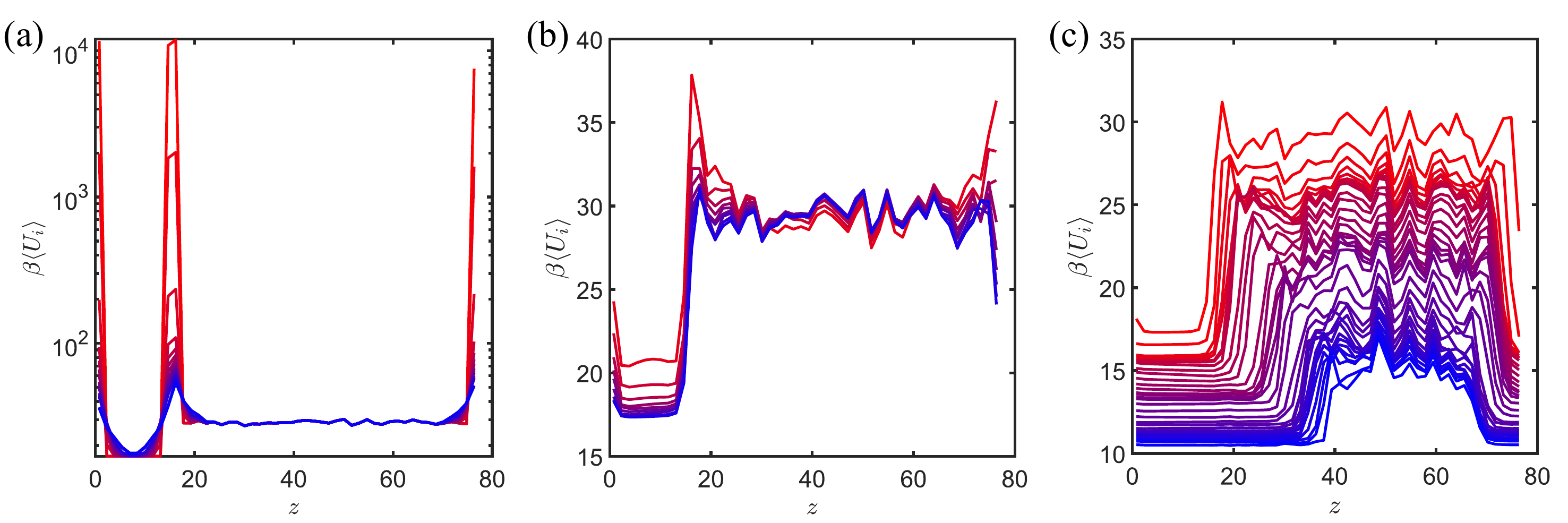}
    \caption{Energy profile over different ranges of iterations, where curves transition from red to blue for early to late iterations within the specified range. (a) Iterations 1 to 30 in 2 iteration steps, (b) 201 to 901 in 100 iteration steps, (c) 1000 to 69000 in 2000 iteration steps.}
    \label{fig:iter}
\end{figure}

This is further evidenced by looking at how the distribution of energy in the system changes during the steepest descent energy minimization following the shift in $z$. These are shown in Fig.\ref{fig:iter}, for different ranges of iterations while minimizing a configuration at $\phi = 0.5327$, where the flat portion at low $z$ corresponds to the crystal phase. Looking at (a), we can immediately see that the bulk of the interfacial energy is relaxed over the first 30 or so iterations, with further slower relaxation over the first 1000 (b), until the interfacial energy is within the variation expected in the disordered glassy phase. It is only after this has finished that the crystal-glass interface begins to move, as shown in (c), demonstrating that relaxation of the interfacial energy takes place entirely independently.

Thus, we can see that energy localized at the interface does not play a role in the migration of the crystalline front. Importantly, this evidences the independence of the method by which the glass is interfaced with a crystal with what is observed.

\section{Effect of polydispersity on UGX dynamics}
Polydispersity is known to affect the dynamics of crystallizing systems \cite{Martin2003,Schope2007}. We note that UGX states are slightly polydisperse ($<$4\%), and that this may have a direct impact by itself on the crystallization dynamics of UGX states. Thus, we prepared CGX states with the same polydispersity, without the algorithm which homogenizes the volume fraction of individual particles, referred to as reCGX states. Note that previous work \cite{Yanagishima2021} showed that the same exercise carried out for bulk glasses created slightly polydisperse glasses with qualitatively the same dynamics as normal glass states.

\begin{figure}
    \centering\includegraphics[width=8cm]{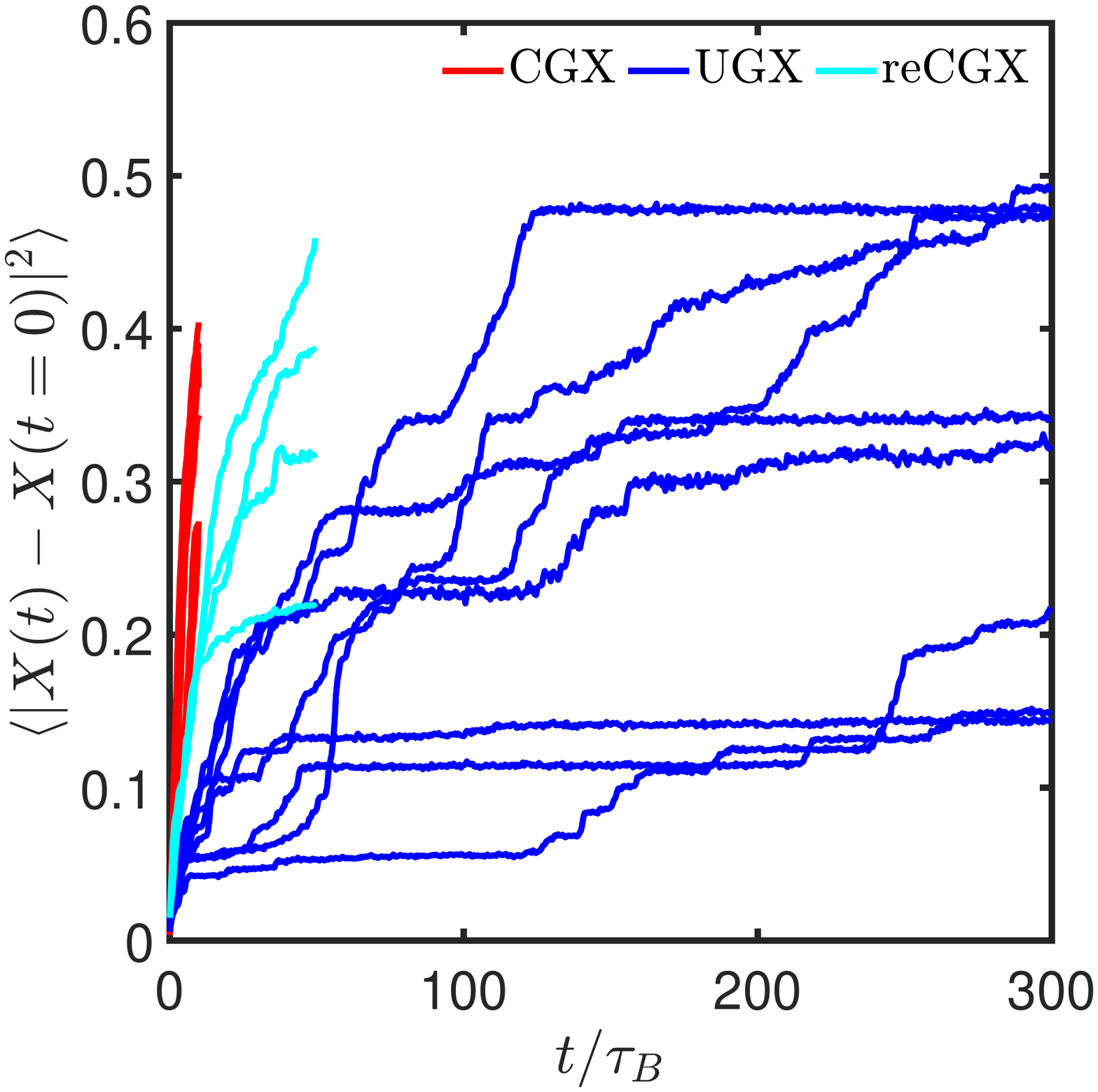}
    \caption{Squared displacements of particles in CGX, UGX and reCGX states at $\phi = 0.517$. Different lines correspond to independent simulations. Note that reCGX states, despite some slowdown, is still qualitatively similar to CGX states.}
    \label{fig:reCGXdst}
\end{figure}

We firstly look at the crystallization dynamics. If the polydispersity is responsible for the slowed crystallization, we should see the same dynamics as UGX states. However, as seen in Fig.\ref{fig:reCGXdst}, it is clear that the dynamics are qualitatively different, with no clear intermittency as well as larger displacements overall. The dominant mechanism supporting the intermittency of the dynamics seems to be unique to UGX states. The migration of the front in reCGX states is consistent and fast enough for us to extract a crystallization speed of $v = 0.83 \pm 0.05$, averaged over eight independent reCGX states. This is slower than in CGX states at the same density, but not significantly so.

We note however that the effect of polydispersity is not limited to the introduction of a slight slowdown. This is reflected in the movement of the crystalline front on finding the initial inherent structure. At $\phi = 0.517$, the front moves by $l_X = 2.3 \pm 0.3$ during energy minimization. This is less than in CGX states where migrations of $l_X>10$ are seen, but still larger than in UGX states, where $l_X = 1.0\pm 0.1$. Since the particles are slightly polydisperse, particle inclusion on the crystal lattice may introduce small lattice distortions, creating barriers in the energy landscape. The fact that they are very low is reflected in the fact that they do not introduce significant intermittency into the dynamics. It is worth noting that the slowdown mechanism here is distinct from in the seminal works on weakly polydisperse systems by Bryant and van Megen \cite{Martin2003,Schope2007}, where slow fractionation may occur which helps crystallization. Particles in our glasses cannot diffuse large enough distances to impact the local particle population; thus, the low barriers are mechanical in nature, and are apparently easily surmounted by thermal fluctuations.

We thus conclude that the unique intermittency in the dynamics seen for UGX states as well as the suppression of crystal front migration are unique to the construction of UG states, and not a product of polydispersity.

\bibliography{biblio}